\documentclass[RNAAS,twocolumn]{aastex63}

\def\src {\mbox{Swift\,J1818.0-1607}}

\usepackage{ulem}
\usepackage{color}

\received{2020 August 2}
\accepted{2020 September 24}
\submitjournal{RNAAS}

\shorttitle{\src}
\shortauthors{Y. Liu and Y.-C. Zou}

\begin{document}

\title{A search for the guest star associated with \src}

\email{yul@hust.edu.cn (YL), zouyc@hust.edu.cn (Y-CZ)}

\author[0000-0002-4421-7282]{Yu Liu}
\affiliation{Department of Astronomy, School of Physics, Huazhong University of Science and Technology, Wuhan 430074, China}

\author[0000-0002-5400-3261]{Yuan-Chuan Zou}
\affiliation{Department of Astronomy, School of Physics, Huazhong University of Science and Technology, Wuhan 430074, China}

\begin{abstract}
We searched the possible historical records for the young magnetar \src, and found a guest star in AD 1798 that might be associated with it.
\end{abstract}

\section{Introduction}

A magnetar is a neutron star with ultra-strong magnetic fields ($B \sim 10^{14}-10^{15} \mathrm{G}$) \citep{Kaspi2017}. Recently found magnetar \src, with the characteristic age \mbox{$\tau_c=P/(2\dot{P})\simeq240$\,yr} in the Milky Way \citep{esposito2020}, is the youngest magnetar discovered to date. Like other relatively young magnetars, this newly identified magnetar offers the possibility to search for its historical guest star \citep{Kaspi2001}.

Magnetars are thought to be born in core-collapse events with high-angular momentum progenitors or in the aftermath of binary NS mergers \citep{Casey2019}. The extra energy injection from magnetars will make these magnetar-powered transients brighter than the normal ones \citep{Zhang2018,Mereghetti2008}. 
If the true age of \src\ is similar to its characteristic age, i.e. $\sim 240$ yr, we might expect that such a bright transient source was witnessed by the ancients.

\section{Historical Records}

With an estimated age of 240 yr, we searched ancient Chinese records during the eighteenth century, which can be found in \citet{BAO1988}, page 380, and other related chapters. We found three closest records as shown in figure \ref{fig:guest}, and translated them as follows:

\begin{figure}[ht!]
    \centering
    \includegraphics[width=0.4\textwidth]{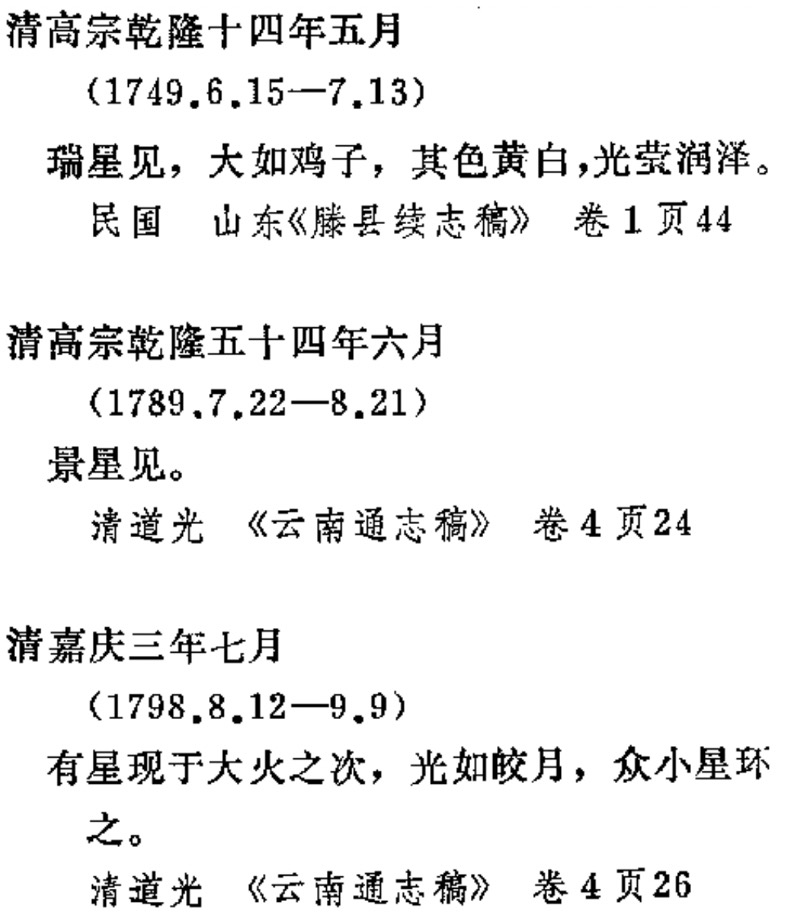}
    \caption{The three records of guest stars during the eighteenth century from \citet{BAO1988},  which were transcribed from the historical books. The first one was from Tengxian Xuzhi Gao (a draft for the history of the county Tengxian), volume 1, page 44. The last two were from Yunnan Tongzhi Gao (a draft for the general history of province Yunnan), volume 2, pages 24 and 26 respectively. The referred records are  shown in figures \ref{fig:AD1749}-\ref{fig:AD1798}.
    \label{fig:guest}}
\end{figure}

\begin{figure}[ht!]
    \centering
    \includegraphics[width=0.4\textwidth]{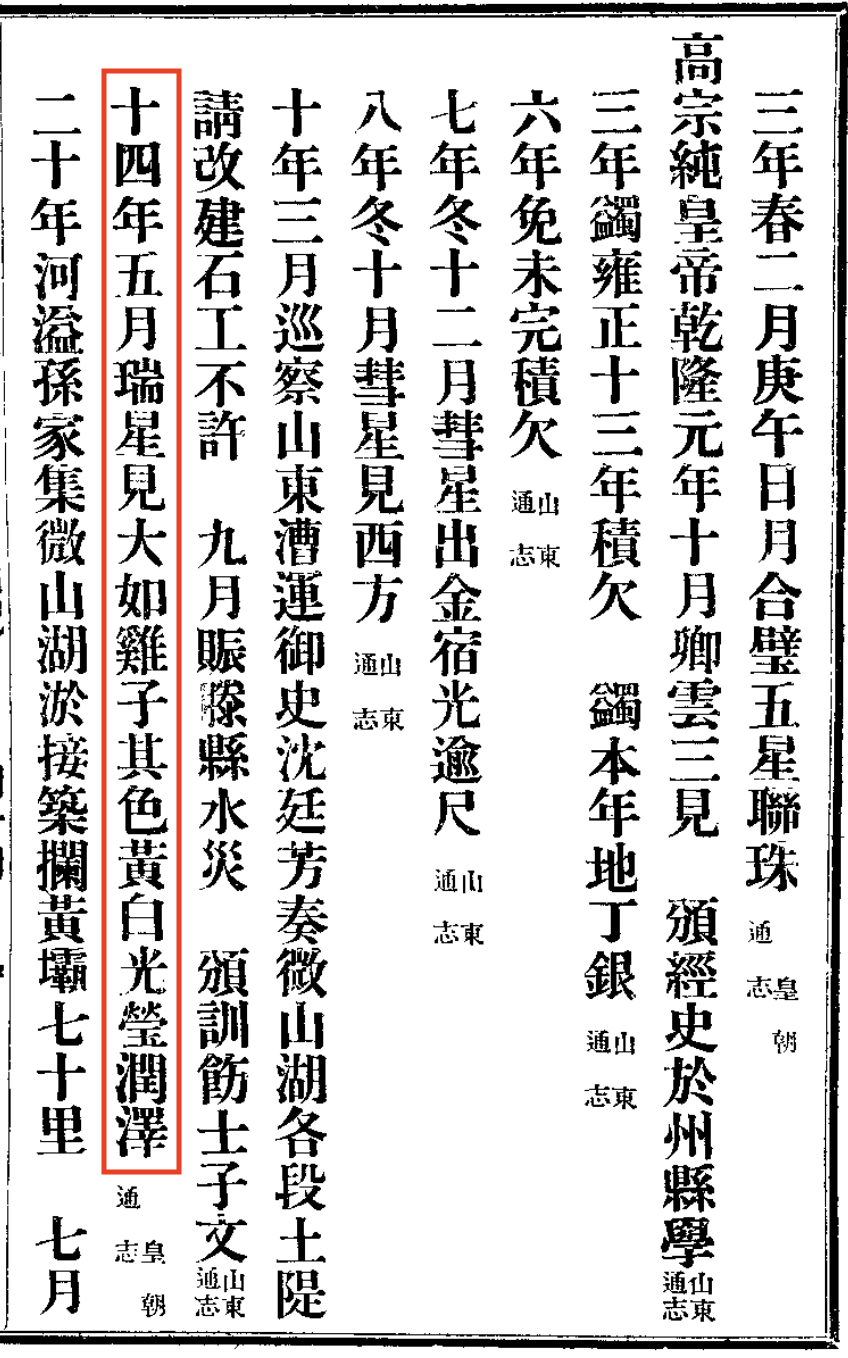}
    \caption{The original record of the guest star in AD 1749, which is from Tengxian Xuzhi Gao, vol. 1, gage 44. \label{fig:AD1749}}
\end{figure}

\begin{figure}[ht!]
    \centering
    \includegraphics[width=0.4\textwidth]{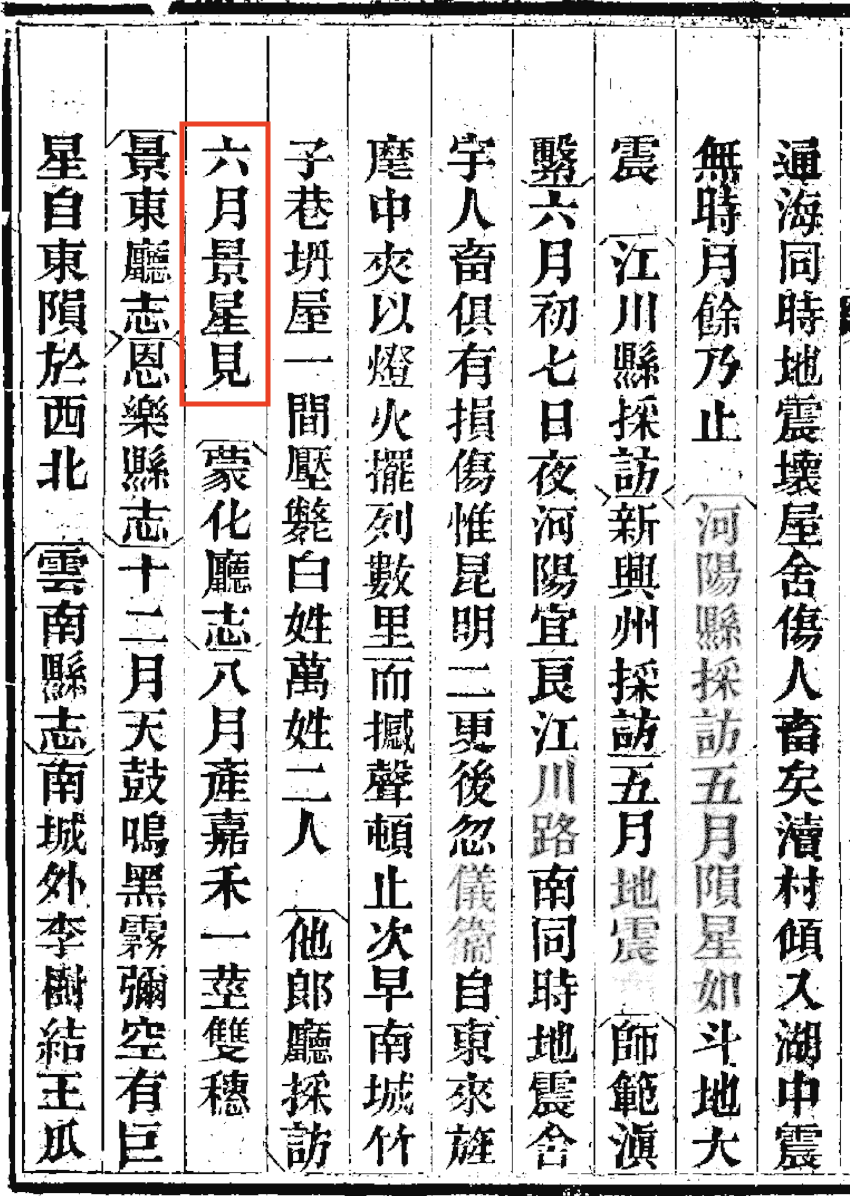}
    \caption{The original record of the guest star in AD 1789, which is from Yunnan Tongzhi Gao, vol. 4, page 24. \label{fig:AD1789}}
\end{figure}

\begin{figure}[ht!]
    \centering
    \includegraphics[width=0.4\textwidth]{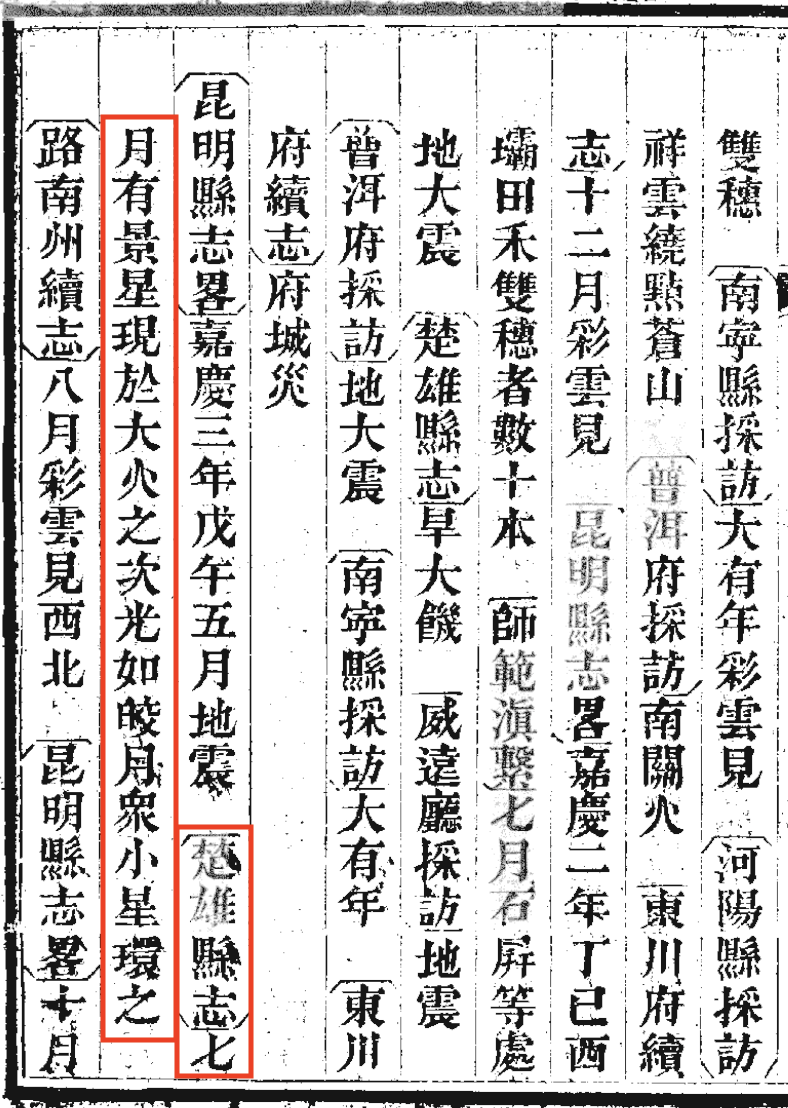}
    \caption{The original record of the guest star in AD 1798, which is from Yunnan Tongzhi Gao, vol. 4, page 26. \label{fig:AD1798}}
\end{figure}

\begin{itemize}
    \item ``During the fifth lunar month of the fourteenth year of Qing Gaozong Qianlong reign period [1749 June 15th-July 13th], a guest star (Rui-Xing) was seen. It was as large as an egg, yellowy white in color and lustre is smooth.'' [Tengxian Xuzhi Gao, vol. 1, gage 44] 
    \item ``During the sixth lunar month of the Fifty-fourth year of Qing Gaozong Qianlong reign period [1789 July 22th-August 21th], a guest star (Jing-Xing) was seen.'' [Yunnan Tongzhi Gao, vol. 4, page 24]
    \item ``During the seventh lunar month of the third year of Qing Jiaqing reign period [1798 August 12th-September 9th], a guest star (Jing-Xing) appeared in Chinese Zodiac of Antares (Dahuo Zhi Ci). It was as bright as the full Moon. Many small stars surrounded it.'' [Yunnan Tongzhi Gao, vol. 4, page 26]
\end{itemize}
In the records, the ``Jing-Xing'' is a kind of ``Rui-Xing'' that normally refers to a star of good event symbol, and the size of an egg corresponds to 3rd or 4th magnitude \citep{Yang2005}.
According to \citet{BAO1988}, the earliest record was in 1690, and the latest one was in 1861. They are unlikely to be associated with a 240 yr olds magnetar. Therefore, those three records shown above should be the most relevant ones. The original records are also shown in figures \ref{fig:AD1749}-\ref{fig:AD1798}.

Two of them, records in AD 1749 and AD 1789, have no location information and can not be further verified. Only the one in AD 1798 attracted our attention. However, the literal meaning of ``Dahuo Zhi Ci'' in the historical record of AD 1798 is not absolutely clear. On the one hand, Antares (Dahuo) is the brightest object in the constellation of Scorpius, and the ancient Chinese astronomers had a custom of describing the visual position of suspected new stars by their nearby brightest stars or asterisms \citep{Wang2002}. On the other hand, Dahuo is also one of the names of astrological signs in the Chinese Zodiac (Xing Ci)\footnote{Oriental astronomers also divided the zodiac into 12 sections. We simply call these 12 divisions as 12 Chinese Zodiac.}, which corresponds to the Zodiac of Antares.  Due to the appearance of both Dahuo and Ci, we suggest ``Dahuo Zhi Ci'' means the Scorpio (Zodiac of Antares). 
In either case, the guest star in AD 1798 is not far from the star Dahuo (Antares, Alpha Scorpii).

\begin{figure}[ht!]
    \centering
    \includegraphics[width=0.4\textwidth]{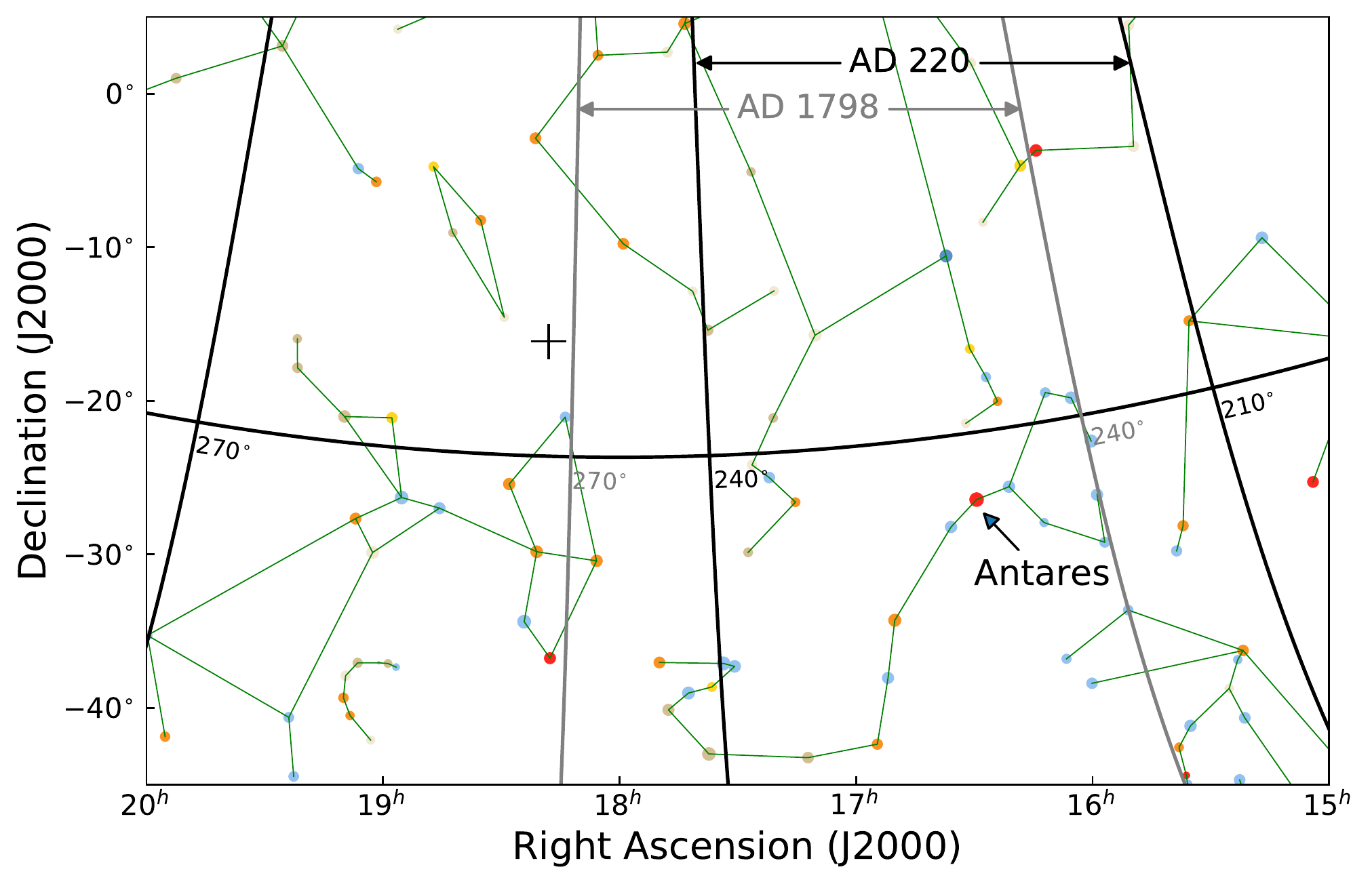}
    \caption{The chart covers two parts of the Chinese Zodiac. \src\ is denoted with $+$ symbol. Antares is the brightest object in this region. Modern constellations are connected by green solid lines. 
    The solid black lines and solid gray lines represent the coordinate system of the ecliptic at epoch AD 220 and AD 1798, respectively, i.e., the regions of Scorpio in these two periods. \label{fig:constellation}}
\end{figure}

Figure \ref{fig:constellation} shows the most probable position of AD 1798 and some nearby modern constellations \citep{Xu2011}, as well as the most probable position of the magnetar \src. 
Due to the precession of the equinox, there are two zodiacs, the tropical or moving zodiac, which is measured from the tropical, and the fixed or sidereal zodiac, which is measured from the fixed star \citep{Gleadow1969}. In our case, the epoch corresponds to AD 1798 or AD 220, respectively.
It is not clear to the authors which epoch should be used. We use different colors to distinguish them.
It is shown that \src\ is not exactly inside the Scorpio. However, stars in Antares were brighter than those in the next zodiac sign Kaus Australis (Ximu). It is quite likely that Antares was used as a reference. Although the visual position of \src\ ($\rm RA=18^h18^m00\fs16$, $\rm Dec= -16^\circ07'53\farcs2$, J2000.0) is within the other Chinese Zodiac Kaus Australis rather than Antares, the spatial coincidence between the guest star and \src\ is still very likely.

\section{Discussion}

If \src\ is indeed associated with the guest star in AD 1798, the age is accurately determined. We can yield an estimate of the initial spin period $P_{0} \sim 375 \rm \ ms$ \citep{Manchester1977,Kaspi2001}, which is heavily based on the assumption that the spindown of \src\ is dominated by magnetic dipole radiation with braking index $n = 3$. Considering the event rate and birth spin period, \src\ seems more likely to come from supernova, although we cannot totally rule out the possibility of a binary NS merger.
As the record says ``Many small stars surrounded it'', it might be an SN 1987A like supernova appeared in the Milky Way.

\bibliography{ref}{}

\begin{thebibliography}{}
\expandafter\ifx\csname natexlab\endcsname\relax\def\natexlab#1{#1}\fi
\providecommand{\url}[1]{\href{#1}{#1}}
\providecommand{\dodoi}[1]{doi:~\href{http://doi.org/#1}{\nolinkurl{#1}}}
\providecommand{\doeprint}[1]{\href{http://ascl.net/#1}{\nolinkurl{http://ascl.net/#1}}}
\providecommand{\doarXiv}[1]{\href{https://arxiv.org/abs/#1}{\nolinkurl{https://arxiv.org/abs/#1}}}

\bibitem[{Esposito {et~al.}(2020)Esposito, Rea, Borghese, Zelati, Viganò,
  Israel, Tiengo, Ridolfi, Possenti, Burgay, Götz, Pintore, Stella, Dehman,
  Ronchi, Campana, Garcia-Garcia, Graber, Mereghetti, Perna, Castillo, Turolla,
  \& Zane}]{esposito2020}
Esposito, P., Rea, N., Borghese, A., {et~al.} 2020, A very young radio-loud
  magnetar.
\newblock \doarXiv{2004.04083}

\bibitem[{{Gleadow}(1969)}]{Gleadow1969}
{Gleadow}, R. 1969, {The origin of the zodiac.}

\bibitem[{{Kaspi} \& {Beloborodov}(2017)}]{Kaspi2017}
{Kaspi}, V.~M., \& {Beloborodov}, A.~M. 2017, \araa, 55, 261,
  \dodoi{10.1146/annurev-astro-081915-023329}

\bibitem[{{Kaspi} {et~al.}(2001){Kaspi}, {Roberts}, {Vasisht}, {Gotthelf},
  {Pivovaroff}, \& {Kawai}}]{Kaspi2001}
{Kaspi}, V.~M., {Roberts}, M.~E., {Vasisht}, G., {et~al.} 2001, \apj, 560, 371,
  \dodoi{10.1086/322515}

\bibitem[{{Law} {et~al.}(2019){Law}, {Margalit}, {Palliyaguru}, {Metzger},
  {Sironi}, {Zheng}, {Berger}, {Margutti}, {Beloborodov}, {Nicholl},
  {Eftekhari}, {Vurm}, \& {Williams}}]{Casey2019}
{Law}, C., {Margalit}, B., {Palliyaguru}, N.~T., {et~al.} 2019, \baas, 51, 319.
\newblock \doarXiv{1903.04691}

\bibitem[{{Manchester} \& {Taylor}(1977)}]{Manchester1977}
{Manchester}, R.~N., \& {Taylor}, J.~H. 1977, {Pulsars}

\bibitem[{{Mereghetti}(2008)}]{Mereghetti2008}
{Mereghetti}, S. 2008, \aapr, 15, 225, \dodoi{10.1007/s00159-008-0011-z}

\bibitem[{{Wang} {et~al.}(2002){Wang}, {Li}, \& {Zhao}}]{Wang2002}
{Wang}, Z., {Li}, Z., \& {Zhao}, Y. 2002, \apjl, 569, L43,
  \dodoi{10.1086/340456}

\bibitem[{{Xu}(2011)}]{Xu2011}
{Xu}, J. 2011, in IAU Symposium, Vol. 260, The Role of Astronomy in Society and
  Culture, ed. D.~{Valls-Gabaud} \& A.~{Boksenberg}, 107--115,
  \dodoi{10.1017/S174392131100319X}

\bibitem[{{Yang} {et~al.}(2005){Yang}, {Park}, {Cho}, \& {Park}}]{Yang2005}
{Yang}, H.-J., {Park}, M.-G., {Cho}, S.-H., \& {Park}, C. 2005, \aap, 435, 207,
  \dodoi{10.1051/0004-6361:20042455}

\bibitem[{{Zhang}(2018)}]{Zhang2018}
{Zhang}, B. 2018, {The Physics of Gamma-Ray Bursts},
  \dodoi{10.1017/9781139226530}

\bibitem[{Zhuang \& Wang(1988)}]{BAO1988}
Zhuang, W., \& Wang, L. 1988, Sylloge of Ancient Chinese Celestial Records
  (Nanjing: Jiangsu Science \& Technology Press).
\newblock \url{https://books.google.ca/books?id=VZC9AAAAIAAJ}

\end{thebibliography}
\bibliographystyle{aasjournal}

\end{document}